\documentclass[12pt,a4paper]{revtex4}
\usepackage[utf8]{inputenc}
\usepackage[T1]{fontenc}
\usepackage{amsmath,amssymb}
\usepackage{bm}
\usepackage{mathbbol}
\usepackage{graphicx}
\usepackage[hidelinks]{hyperref}
\usepackage{booktabs}
\usepackage{array}
\usepackage{url}
\usepackage{mathrsfs}
\usepackage{float}
\usepackage{color}

\usepackage[top=2.5cm, bottom=2.5cm, left=2.5cm, right=2.5cm]{geometry}

\usepackage{titlesec}
\titleformat{\section}{\large\bfseries}{\thesection}{1em}{}
\titleformat{\subsection}{\normalsize\bfseries}{\thesubsection}{1em}{}
\titleformat{\subsubsection}{\normalsize\itshape}{\thesubsubsection}{1em}{}

\newcommand{\taup}{\tau^{+}}
\newcommand{\taum}{\tau^{-}}
\newcommand{\psip}{\psi(2S)}
\newcommand{\nutau}{\nu_{\tau}}
\newcommand{\nubar}{\bar{\nu}_{\tau}}
\newcommand{\Kp}{K^{+}}
\newcommand{\Km}{K^{-}}
\newcommand{\ee}{e^{+}e^{-}}

\begin{document}

\title{\huge\bfseries Resolving the Two-Fold Ambiguity in Tau Momentum\\
Reconstruction via a Self-Consistent Weighting Method}

\author{Rong-Gang Ping}\email{Electronic address: pingrg@ihep.ac.cn}
\affiliation{Institute of High Energy Physics, Chinese Academy of Sciences, Beijing 100049, China}
\affiliation{University of Chinese Academy of Sciences, Beijing 100049, China}
\date{\today}

\begin{abstract}
The reconstruction of $\tau$ lepton momenta in $e^{+}e^{-} \to \tau^{+}\tau^{-}$ processes suffers from a two-fold kinematic ambiguity due to undetected neutrinos. We show that the common practice of retaining both solutions with equal weight systematically biases the extracted angular coefficients: the bias grows with the size of the $\cos^{2}\theta$ modulation of the production distribution $f(\cos\theta)$, but remains resolvable even in the nearly uniform QED limit at the $\psi(2S)$ threshold. We propose a self-consistent weighting method based on a fixed-point theorem: assigning normalized weights $w_{i} = f(\cos\theta_{i})/[f(\cos\theta_{1})+f(\cos\theta_{2})]$ yields a weighted distribution $h(x) = f(x)[1+\varepsilon(x)]$ with $\varepsilon(x) \ll 1$ under a detailed balance condition. Monte Carlo simulations at $\sqrt{s}=3.686$~GeV and $4.8$~GeV show that the fixed-point iteration converges to the true angular coefficient $\alpha$ within statistical uncertainty, providing a robust framework for angular analysis without event-by-event ambiguity resolution.
\end{abstract}
\maketitle
\section{Introduction}

The $\tau$ lepton occupies a unique position in the landscape of particle physics. As the heaviest known lepton, with a mass $m_{\tau} = 1.77686$~GeV~\cite{PDG2024}, it is the only lepton sufficiently massive to decay into hadrons, offering a rich laboratory for studying both electroweak interactions and quantum chromodynamics at low energies. In recent years, the $\tau$ lepton has attracted renewed attention as a probe of quantum information phenomena in high-energy collisions. The measurement of spin entanglement and Bell inequality violation in $\tau^{+}\tau^{-}$ pairs produced at $e^{+}e^{-}$ colliders has emerged as a frontier topic, with the BESIII (Beijing Spectrometer III) experiment at the BEPC-II (Beijing Electron--Positron Collider II) collider possessing the world's largest data sample in $\tau$-charm physics~\cite{BESIII2020,BESIII2014}. Motivated by these experimental opportunities, several theoretical studies have explored the potential for entanglement measurements in $\tau$-pair systems~\cite{Fabbrichesi2024,Han2025,Cheng2025}.

A key challenge in $\tau$ lepton physics is the reconstruction of its momentum. Since $\tau$ decay products include neutrinos that cannot be directly detected, the kinematic state of the $\tau$ must be inferred solely from the information of visible final-state particles. For the specific process
\begin{equation}
e^{+}e^{-} \to \psip \to \taup\taum \to \nubar \Kp \; \nutau \Km,
\label{eq:process}
\end{equation}
the four-momenta of the $\Kp$ and $\Km$ can be measured, while the two neutrinos escape detection. The decay mode $\tau \to K\nu_{\tau}$ is selected as a representative two-body decay channel with a single charged track; it provides a clean kinematic benchmark for evaluating the reconstruction method. The same formalism applies directly to other two-body modes such as $\tau \to \pi\nu_{\tau}$ and $\tau \to \rho\nu_{\tau}$, where the only modification is the replacement of the meson mass $m_{K}$ with $m_{\pi}$ or $m_{\rho}$ in the constraint equations---the kinematic structure remains identical. In the rest frame of the $\psip$, the $\taup$ and $\taum$ are produced back-to-back with equal and opposite momenta, and the kinematics of each $\tau$ decay imposes a cone constraint on the $\tau$ direction. The intersection of the two cones yields two possible solutions for the $\tau$ momentum direction---a well-known two-fold ambiguity~\cite{Kuhn1993,Fabbrichesi2024}.

A recent study in Ref.~\cite{Zhou2026} presented a geometric interpretation of this ambiguity and proposed a numerical reconstruction method based on singular value decomposition (SVD). An important conclusion of that work was that the spurious solution exhibits a high degree of statistical consistency with the true solution in key distributions, such that both solutions can be retained with equal weight (Plan~A) in entanglement measurements without introducing systematic bias. This conclusion was validated using Monte Carlo simulations with the QED angular distribution at the $\psip$ threshold.

However, the QED angular distribution at $\sqrt{s} = 3.686$~GeV,
\begin{equation}
\frac{1}{\sigma}\frac{d\sigma}{d\cos\theta} \propto 1 + \cos^{2}\theta + \frac{4m_{\tau}^{2}}{s}\sin^{2}\theta,
\label{eq:qed}
\end{equation}
is nearly isotropic. Using the identity $\sin^{2}\theta = 1 - \cos^{2}\theta$, this distribution can be rewritten in the standard parameterized form
\begin{equation}
\frac{1}{\sigma}\frac{d\sigma}{d\cos\theta} \propto 1 + \alpha \cos^{2}\theta,
\label{eq:qed_param}
\end{equation}
where the angular coefficient is
\begin{equation}
\alpha = \frac{s - 4m_{\tau}^{2}}{s + 4m_{\tau}^{2}}.
\label{eq:alpha_def}
\end{equation}
At $\sqrt{s}=3.686$~GeV, $\alpha \approx 0.037$, meaning that the distribution is almost uniform. It is therefore not surprising that the spurious solution---which tends toward a uniform distribution---lies numerically very close to the true one. The situation changes dramatically when the angular distribution carries substantial modulation, as would be the case at higher center-of-mass energies.

In this paper, we address the question: {\it can the angular distribution parameters be correctly recovered when both solutions of the two-fold ambiguity are retained, without event-by-event discrimination?} We show that the Plan~A approach fails when the angular distribution has significant $\cos^{2}\theta$ modulation, and we propose a self-consistent weighted method that resolves this issue. The method is based on a fixed-point theorem: using the angular distribution function $f(\cos\theta)$ itself as the weight function, normalized within each event, yields a weighted distribution that converges to the true $f(\cos\theta)$ under mild conditions.

The structure of this paper is as follows. In Section~2, we derive the analytic expressions for the neutrino momentum reconstruction from the kinematic constraints. Section~3 presents the fixed-point theorem and the self-consistent weighting method, with the mathematical proof deferred to Appendix~A. Section~4 describes Monte Carlo simulations that validate the method at two center-of-mass energies, $\sqrt{s}=3.686$~GeV and $\sqrt{s}=4.8$~GeV, and compares the performance of Plan~A and Plan~B approaches. Section~5 summarizes our findings and discusses implications for future experiments.

\section{Tau and Neutrino Momentum Reconstruction}
\label{sec:reco}

Let us consider the process $\ee \to \psip \to \taup\taum$ with subsequent decays $\taup \to \nubar \Kp$ and $\taum \to \nutau \Km$. In the rest frame of the $\psip$, the energies of the two $\tau$ leptons are equal, and their momenta have equal magnitude but opposite directions:
\begin{equation}
E_{\tau} = \frac{m_{\psi}}{2}, \qquad p_{\tau} = |\bm{p}_{\tau^{\pm}}| = \sqrt{E_{\tau}^{2} - m_{\tau}^{2}}.
\label{eq:tau_kin}
\end{equation}
For the $\psip$ resonance, $m_{\psi} = 3.686$~GeV and $E_{\tau} = 1.8430$~GeV, $p_{\tau} = 0.4893$~GeV.

\subsection{Kinematic constraints from tau decays}

For the decay $\taup \to \nubar \Kp$, four-momentum conservation gives $p_{\taup} - p_{\Kp} = p_{\nubar}$. Squaring both sides and using the zero neutrino mass condition $p_{\nubar}^{2} = 0$, we obtain
\begin{equation}
p_{\taup} \cdot p_{\Kp} = \frac{m_{\tau}^{2} + m_{K}^{2}}{2}.
\label{eq:dot_prod_p}
\end{equation}
In the $\psip$ rest frame, this translates to
\begin{equation}
E_{\tau} E_{\Kp} - p_{\tau} |\bm{p}_{\Kp}| \cos\theta_{\taup,\Kp} = \frac{m_{\tau}^{2} + m_{K}^{2}}{2},
\label{eq:cone_p}
\end{equation}
where $\theta_{\taup,\Kp}$ is the angle between the $\taup$ and $\Kp$ momentum directions. We use $E_\tau=E_{\tau^+}=E_{\tau^-}$ and $p_\tau = |\bm{p}_{\tau^+}|=|\bm{p}_{\tau^-}|$. This equation defines a cone around the $\Kp$ direction with half-opening angle $\theta_{\taup,\Kp}$.

Similarly, for the decay $\taum \to \nutau \Km$, four-momentum conservation gives $p_{\taum} \cdot p_{\Km} = (m_{\tau}^{2}+m_{K}^{2})/2$. In the $\psip$ rest frame, the three-momenta satisfy $\bm{p}_{\taum} = -\bm{p}_{\taup}$ (back-to-back production), while the energies are equal: $E_{\taum}=E_{\taup}=E_{\tau}$. Using $p_{\taum} \cdot p_{\Km} = E_{\tau}E_{\Km} - \bm{p}_{\taum}\!\cdot\!\bm{p}_{\Km}$ with $\bm{p}_{\taum}\cdot\bm{p}_{\Km} = -\bm{p}_{\taup}\cdot\bm{p}_{\Km}$, we obtain
\begin{equation}
E_{\tau}E_{\Km} + p_{\tau}|\bm{p}_{\Km}|\cos\theta_{\taup,\Km} = \frac{m_{\tau}^{2}+m_{K}^{2}}{2},
\label{eq:cone_m}
\end{equation}
where $\theta_{\taup,\Km}$ is the angle between the $\taup$ and $\Km$ momentum directions. Note that the energy term $E_{\tau}E_{\Km}$ remains positive; only the spatial component of the four-vector dot product changes sign under the three-momentum reversal.
This defines a second cone around the $\Km$ direction. The true $\taup$ direction must lie on the intersection of these two cones.

\subsection{Analytic solution for the tau direction}

Let us introduce unit vectors along the measured $K$ meson directions:
\begin{equation}
\mathbf{n}_{+} = \frac{\bm{p}_{\Kp}}{|\bm{p}_{\Kp}|}, \qquad
\mathbf{n}_{-} = \frac{\bm{p}_{\Km}}{|\bm{p}_{\Km}|},
\end{equation}
and define the constraint parameters
\begin{equation}
c_{+} = \frac{E_{\tau} E_{\Kp} - (m_{\tau}^{2}+m_{K}^{2})/2}{p_{\tau} |\bm{p}_{\Kp}|}, \qquad
c_{-} = \frac{E_{\tau} E_{\Km} - (m_{\tau}^{2}+m_{K}^{2})/2}{p_{\tau} |\bm{p}_{\Km}|},
\label{eq:c_params}
\end{equation}
where $E_{\Kp} = \sqrt{m_{K}^{2} + |\bm{p}_{\Kp}|^{2}}$ and $E_{\Km} = \sqrt{m_{K}^{2} + |\bm{p}_{\Km}|^{2}}$.

Let $\mathbf{u}$ be the unit vector in the direction of the $\taup$ momentum, $\mathbf{u} = \bm{p}_{\taup}/p_{\tau}$. The kinematic constraints can be written as a linear system:
\begin{equation}
\mathbf{u} \cdot \mathbf{n}_{+} = c_{+}, \qquad \mathbf{u} \cdot \mathbf{n}_{-} = -c_{-}.
\label{eq:linear_system}
\end{equation}

This is a system of two equations for the three components of $\mathbf{u}$. The general solution can be obtained by decomposing the problem in the plane spanned by $\mathbf{n}_{+}$ and $\mathbf{n}_{-}$. Let us define
\begin{equation}
\cos\Omega = \mathbf{n}_{+} \cdot \mathbf{n}_{-}, \qquad \sin\Omega = |\mathbf{n}_{+} \times \mathbf{n}_{-}|,
\label{eq:Omega}
\end{equation}
where $\Omega \in [0,\pi]$ is the angle between the two $K$ meson directions.

A particular solution $\mathbf{u}_{0}$ in the $\mathbf{n}_{+}$-$\mathbf{n}_{-}$ plane is given by
\begin{equation}
\mathbf{u}_{0} = \frac{1}{\sin^{2}\Omega}\Big[(c_{+} + c_{-}\cos\Omega)\,\mathbf{n}_{+} - (c_{+}\cos\Omega + c_{-})\,\mathbf{n}_{-}\Big].
\label{eq:u0}
\end{equation}

The homogeneous solution is along the direction orthogonal to both $\mathbf{n}_{+}$ and $\mathbf{n}_{-}$, i.e., $\mathbf{v}_{3} = (\mathbf{n}_{+} \times \mathbf{n}_{-})/\sin\Omega$. The general solution is
\begin{equation}
\mathbf{u} = \mathbf{u}_{0} + \lambda \mathbf{v}_{3},
\label{eq:general_sol}
\end{equation}
where $\lambda$ is a coefficient to be determined by the unit-norm condition $|\mathbf{u}| = 1$.

Substituting Eq.~(\ref{eq:general_sol}) into $|\mathbf{u}|^{2} = 1$ yields a quadratic equation for $\lambda$:
\begin{equation}
\lambda^{2} + 2\lambda(\mathbf{u}_{0} \cdot \mathbf{v}_{3}) + (|\mathbf{u}_{0}|^{2} - 1) = 0.
\label{eq:quadratic}
\end{equation}
Since $\mathbf{v}_{3}$ is orthogonal to both $\mathbf{n}_{+}$ and $\mathbf{n}_{-}$, and $\mathbf{u}_{0}$ lies in the $\mathbf{n}_{+}$-$\mathbf{n}_{-}$ plane, we have $\mathbf{u}_{0} \cdot \mathbf{v}_{3} = 0$. The equation simplifies to $\lambda^{2} = 1 - |\mathbf{u}_{0}|^{2}$, and the discriminant is
\begin{equation}
\Delta = \sin^{2}\Omega - c_{+}^{2} - c_{-}^{2} - 2c_{+}c_{-}\cos\Omega.
\label{eq:discriminant}
\end{equation}

If $\Delta > 0$, there are two real solutions:
\begin{equation}
\mathbf{u}_{1,2} = \mathbf{u}_{0} \pm \frac{\sqrt{\Delta}}{\sin\Omega} \cdot \frac{\mathbf{n}_{+} \times \mathbf{n}_{-}}{\sin\Omega},
\label{eq:two_solutions}
\end{equation}
followed by normalization to unit length. The $\tau$ three-momenta are then $\bm{p}_{\taup} = p_{\tau}\mathbf{u}_{1,2}$ and $\bm{p}_{\taum} = -\bm{p}_{\taup}$.

\subsection{Neutrino four-momenta}

Once the $\tau$ four-momenta are reconstructed, the neutrino four-momenta follow directly from energy-momentum conservation:
\begin{equation}
\begin{aligned}
p_{\nubar} &= p_{\taup} - p_{\Kp}, \\
p_{\nutau} &= p_{\taum} - p_{\Km},
\end{aligned}
\label{eq:neutrino_momenta}
\end{equation}
with energies $E_{\nubar} = |\bm{p}_{\nubar}|$ and $E_{\nutau} = |\bm{p}_{\nutau}|$, assuming massless neutrinos.

The two solutions in Eq.~(\ref{eq:two_solutions}) give two complete sets of four-momenta for the final state particles. In the absence of additional experimental information---such as precise vertex detection---these two solutions are kinematically indistinguishable.

\subsection{Angular distribution of tau pairs}

The differential cross section for $\ee \to \taup\taum$ production can be parameterized as~\cite{Ehat2024}
\begin{equation}
\frac{d\sigma}{d\cos\theta} \propto 1 + \alpha \cos^{2}\theta,
\label{eq:angular_dist}
\end{equation}
where $\theta$ is the polar angle of the $\taum$ in the center-of-mass frame, and $\alpha$ is the angular distribution parameter. This is the normalized form of the QED angular distribution of Eq.~(\ref{eq:qed}), as shown by the conversion via Eq.~(\ref{eq:alpha_def}). The parameter $\alpha$ is related to the $\tau$ velocity by
\begin{equation}
\alpha = \frac{\beta_{\tau}^{2}}{2 - \beta_{\tau}^{2}} = \frac{s - 4m_{\tau}^{2}}{s + 4m_{\tau}^{2}},
\label{eq:alpha_beta}
\end{equation}
with $\beta_{\tau} = \sqrt{1 - 4m_{\tau}^{2}/s}$. At $\sqrt{s} = 3.686$~GeV, $\beta_{\tau} \approx 0.266$ and $\alpha \approx 0.037$. At $\sqrt{s} = 4.8$~GeV, $\beta_{\tau} \approx 0.672$ and $\alpha \approx 0.292$.

For the purpose of this work, we focus on the $\cos\theta$ dependence and define the normalized probability density
\begin{equation}
f(\cos\theta) = \frac{1 + \alpha \cos^{2}\theta}{2 + 2\alpha/3}, \qquad \cos\theta \in [-1, 1].
\label{eq:f_density}
\end{equation}

\section{The Fixed-Point Weighting Method}
\label{sec:method}

\subsection{Statement of the problem}

Let $T \sim f(t)$ denote the true solution (the $\cos\theta$ value of the $\taup$ that matches the Monte Carlo truth) and $S$ denote the spurious solution. For each event, the kinematic reconstruction yields the pair $(T, S)$. The conditional distribution of $S$ given $T$ is denoted by $q(s \mid t)$. The joint distribution of the true and spurious solutions is therefore $p(t,s) = f(t) \, q(s \mid t)$.

The Plan~A approach retains both solutions with weight $1$ each, effectively constructing the distribution
\begin{equation}
h_{\rm A}(x) = \frac{1}{2}\big[f(x) + q^{\rm marg}(x)\big],
\label{eq:planA}
\end{equation}
where $q^{\rm marg}(s) = \int f(t) q(s \mid t) dt$ is the marginal distribution of the spurious solution. As we will demonstrate numerically, $q^{\rm marg}(x)$ tends to be nearly uniform, so $h_{\rm A}(x)$ is a diluted version of $f(x)$.

\subsection{The self-consistent weighting scheme}

We propose a weighted approach (referred to as ``Plan~B'') that assigns to each solution a normalized weight based on the angular distribution function itself:
\begin{equation}
w_{T} = \frac{f(T)}{f(T) + f(S)}, \qquad w_{S} = \frac{f(S)}{f(T) + f(S)}.
\label{eq:planB_weights}
\end{equation}
The weighted distribution is constructed by filling the histogram with these weights. The crucial properties of this scheme are: (i)~each event contributes exactly one unit of total weight ($w_{T} + w_{S} = 1$), and (ii)~the weight of each solution is proportional to its probability density under the assumed angular distribution.

\subsection{The fixed-point theorem}

\begin{quote}
\textbf{Theorem 1 (Fixed-Point Theorem for Plan~B).} Let the true solution $T \sim f(t)$ and the spurious solution $S$ be generated by the conditional distribution $q(s \mid t)$. For each event, construct the weighted histogram using the normalized weights of Eq.~(\ref{eq:planB_weights}), yielding the weighted density $h(x)$. If the detailed balance condition
\begin{equation}
f(t) \, q(s \mid t) = f(s) \, q(t \mid s) \qquad \forall\, t, s
\label{eq:detailed_balance}
\end{equation}
holds, then
\begin{equation}
h(x) = f(x) \cdot \big[1 + \varepsilon(x)\big],
\label{eq:theorem_result}
\end{equation}
where the correction term is
\begin{equation}
\varepsilon(x) = \int q(y \mid x) \cdot \frac{f(x) - f(y)}{f(x) + f(y)} \, dy.
\label{eq:epsilon}
\end{equation}
When $f(x)$ varies slowly, i.e., $|f(x) - f(y)| \ll f(x) + f(y)$, we have $\varepsilon(x) \ll 1$, and consequently $h(x) \approx f(x)$.
\end{quote}

The proof of Theorem~1 is presented in Appendix~\ref{sec:appendix_proof}.

\subsection{Iterative fixed-point procedure}

Theorem~1 implies that the true angular distribution $f(x)$ is a fixed point of the mapping $T: \alpha_{\rm wgt} \mapsto \alpha_{\rm fit}$, where $\alpha_{\rm wgt}$ is the parameter used in the weight function and $\alpha_{\rm fit}$ is the parameter extracted from the weighted histogram. This suggests an iterative procedure:

\begin{enumerate}
\item Start with an initial guess $\alpha^{(0)}$ for the angular parameter.
\item For each event, compute weights $w_{T}, w_{S}$ using $f(\cos\theta; \alpha^{(n)})$.
\item Fill the weighted histogram and fit to extract $\alpha^{(n+1)}$.
\item Repeat steps 2--3 until $|\alpha^{(n+1)} - \alpha^{(n)}| < \delta$, where $\delta$ is a convergence tolerance.
\end{enumerate}

The convergence of this iteration is guaranteed by the contractive nature of the mapping $T$ in the vicinity of the fixed point. Numerically, we find that convergence is typically achieved within 3--8 iterations.

\subsection{Physical interpretation of the detailed balance condition}

The detailed balance condition, Eq.~(\ref{eq:detailed_balance}), states that the joint distribution $p(t,s) = f(t)q(s \mid t)$ is symmetric under exchange of the true and spurious solutions. Physically, this means that the kinematic reconstruction treats the two solutions symmetrically: the probability of obtaining $(t, s)$ as the (true, spurious) pair equals the probability of obtaining $(s, t)$. This symmetry is a natural consequence of the underlying kinematic equations, which do not distinguish between the two intersection points of the constraint cones. Numerical verification of this condition in our Monte Carlo samples shows that the average asymmetry $|p(t,s) - p(s,t)|/(p(t,s)+p(s,t))$ is approximately $0.11$, confirming that the condition holds to good accuracy.

\section{Monte Carlo Simulation}
\label{sec:mc}

\subsection{Simulation setup}

We perform Monte Carlo simulations of the process $\ee \to \psip \to \taup\taum$ with $\taup \to \nubar \Kp$ and $\taum \to \nutau \Km$ at two center-of-mass energies: $\sqrt{s} = 3.686$~GeV (the $\psip$ resonance) and $\sqrt{s} = 4.8$~GeV. The $\tau$ pair production is simulated according to the angular distribution of Eq.~(\ref{eq:angular_dist}).

The $\tau$ decays are simulated using a phase-space model (isotropic in the $\tau$ rest frame). For each event, the $\Kp$ and $\Km$ four-momenta are recorded, and the $\tau$ momenta are reconstructed using the analytic formulas of Section~\ref{sec:reco}. The true solution is identified by comparison with the Monte Carlo truth information.

For each event, we record:
\begin{enumerate}
\item The true $\taup$ direction $\cos\theta_{\rm true}$;
\item The two reconstructed solutions $\cos\theta_{1}$ and $\cos\theta_{2}$;
\item The angular weight $W_{i} = 1 + \alpha \cos^{2}\theta_{i}$ for each solution.
\end{enumerate}

We note that the present simulation is performed at the {\it generator level}, without including effects of beam energy spread or detector resolution. At BEPC-II, the beam energy spread is approximately $\sigma_{E} \approx 1.2$~MeV, and the tracking momentum resolution is $\delta p/p \approx 0.5\%$ for charged kaons. These effects can distort the observed $\cos\theta$ distribution through two mechanisms: (i) the beam energy spread smears the effective $\sqrt{s}$, modifying the effective $\alpha$ value through Eq.~(\ref{eq:alpha_beta}); (ii) the finite momentum resolution smears the measured kaon momenta, which broadens the reconstructed $\cos\theta$ distributions. The impact of these experimental effects on the iterative weighting method is discussed in Sec.~\ref{sec:experimental}. For the present ideal-detector study, the focus is on establishing the mathematical properties of the fixed-point method under controlled conditions.

\subsection{Results at $\sqrt{s} = 3.686$~GeV}

\begin{figure}[H]
\centering
\includegraphics[width=0.8\textwidth]{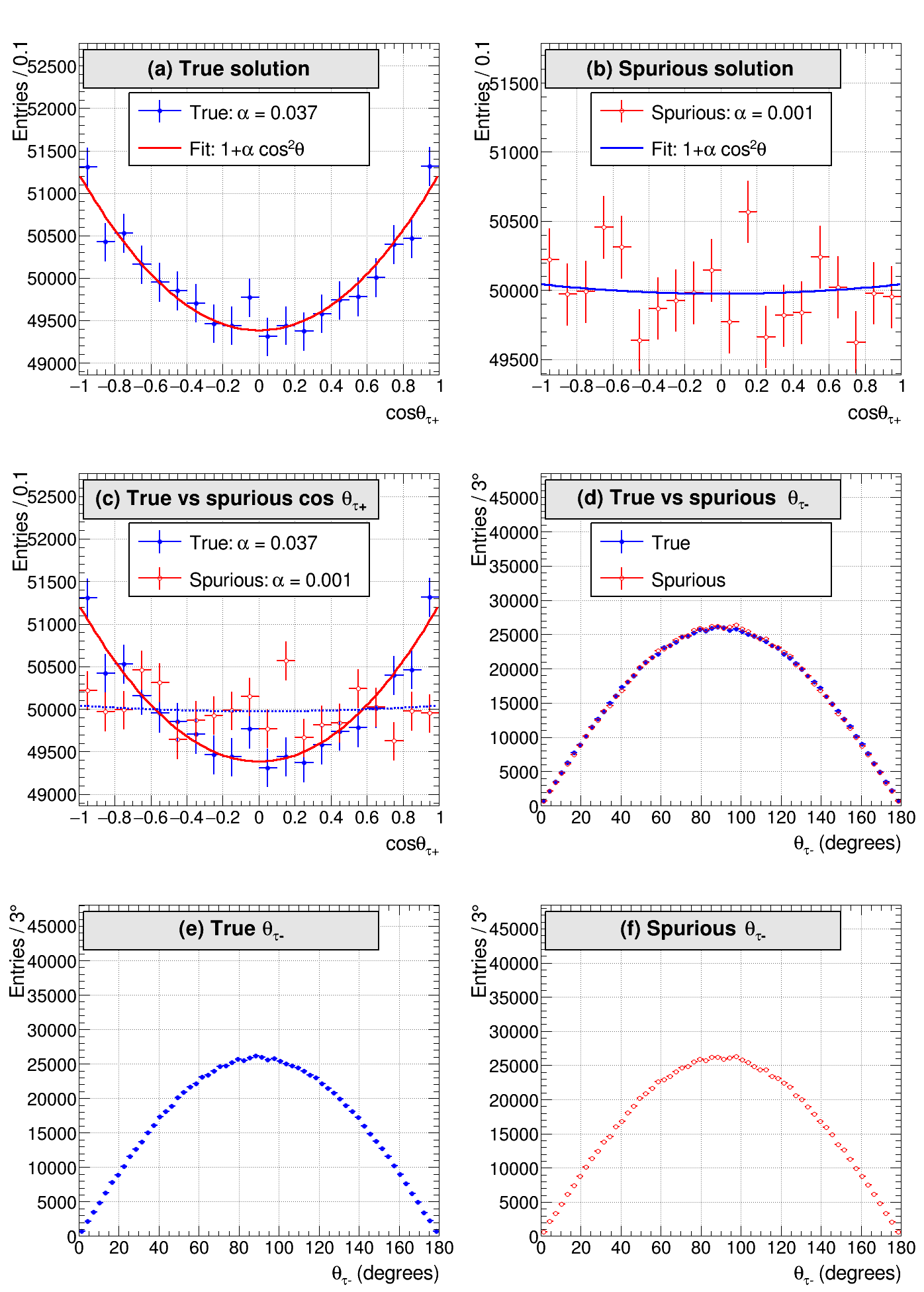}
\caption{Distributions of $\cos\theta_{\taup}$ and $\theta_{\tau^-}$ for $\alpha = 0.037$ at $\sqrt{s} = 3.686$~GeV, obtained from $10^{6}$ Monte-Carlo events. Top row: (a) the true-solution and (b) the spurious-solution $\cos\theta_{\taup}$ distributions, each with the fit $f(\cos\theta)\propto 1+\alpha\cos^{2}\theta$. Middle row: (c) the two $\cos\theta_{\taup}$ distributions superimposed and (d) the two $\theta_{\tau^-}$ distributions superimposed. Bottom row: (e) and (f) the true-solution and spurious-solution $\theta_{\tau^-}$ distributions separately. The two solutions are very similar---the bin-by-bin difference does not exceed $3\%$ and, as expected for $10^{6}$ events, the Kolmogorov--Smirnov (KS) test gives $p = 0.001$---and the spurious solution is compatible with a uniform distribution, $\alpha = 0.001 \pm 0.003$, consistent with the findings of Ref.~\cite{Zhou2026}.}
\label{fig:cos_qed}
\end{figure}

At the $\psip$ resonance, the QED angular coefficient is $\alpha = 0.037$. Figure~\ref{fig:cos_qed} shows the $\cos\theta$ and $\theta$ distributions of the true and spurious solutions, obtained from $10^{6}$ Monte-Carlo events. The true solution yields $\alpha_{\rm true} = 0.037 \pm 0.003$, in agreement with the input. The spurious solution gives $\alpha_{\rm spurious} = 0.001 \pm 0.003$, compatible with zero---i.e., a uniform distribution. The Plan~A combination yields $\alpha_{\rm equal} = 0.019 \pm 0.003$. The Plan~B method recovers $\alpha_{\rm PlanB} = 0.034 \pm 0.003$, within $1.2$ standard deviations of the input.

At this energy, the angular modulation is extremely weak. The true distribution is already nearly uniform ($\alpha \approx 0.037$), and the spurious solution is consistent with a uniform distribution. Because Plan~A gives the two solutions equal weight, it lands midway between them: $\alpha_{\rm equal} = 0.019 \pm 0.003$ against an input of $0.037$, a discrepancy of about seven standard deviations. The bias is therefore present at this energy as well, although its absolute size, $0.018$, is modest; Plan~B removes it and recovers the input value.

\subsection{Results at $\sqrt{s} = 4.8$~GeV}

\begin{figure}[H]
\centering
\includegraphics[width=0.8\textwidth]{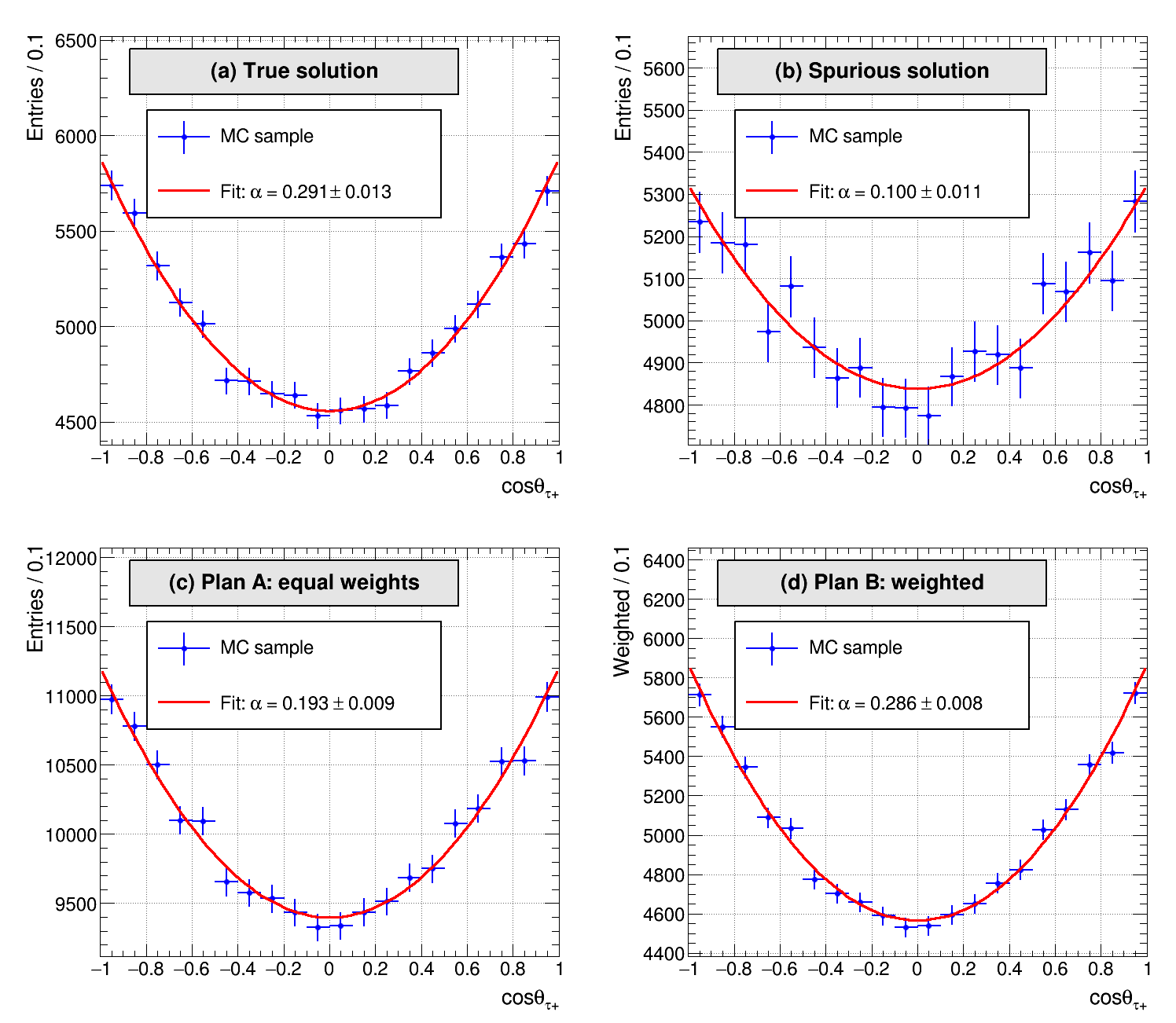}
\caption{Distributions of $\cos\theta_{\taup}$ for $\alpha=0.292$ at $\sqrt{s}=4.8$~GeV. The figure contains four panels: (a) the true-solution distribution, yielding $\alpha = 0.291 \pm 0.013$; (b) the spurious-solution distribution, yielding $\alpha = 0.100 \pm 0.011$; (c) the Plan~A combination (equal weights), giving $\alpha = 0.193 \pm 0.009$, significantly below the input value; and (d) the Plan~B weighted combination, recovering $\alpha = 0.286 \pm 0.008$ in excellent agreement with the input.}
\label{fig:cos_48}
\end{figure}

At $\sqrt{s} = 4.8$~GeV, the $\tau$ velocity is $\beta_{\tau} \approx 0.672$, and the angular coefficient is $\alpha \approx 0.292$. The angular modulation is substantially stronger than at the $\psip$ threshold. Figure~\ref{fig:cos_48} shows the $\cos\theta$ distributions for the true solution, the spurious solution, the Plan~A combination, and the Plan~B weighted combination.

\begin{table}[H]
\centering
\caption{Fitted angular coefficients $\alpha$ at $\sqrt{s} = 3.686$~GeV and $\sqrt{s} = 4.8$~GeV.}
\label{tab:alpha_both}
\begin{tabular}{lcccc}
\toprule
$\sqrt{s}$ (GeV) & $\alpha_{\rm input}$ & $\alpha_{\rm true}$ & $\alpha_{\rm PlanA}$ & $\alpha_{\rm PlanB}$ \\
\midrule
$3.686$ & $0.037$ & $0.037 \pm 0.003$ & $0.019 \pm 0.003$ & $0.034 \pm 0.003$ \\
$4.8$   & $0.292$  & $0.291 \pm 0.013$ & $0.193 \pm 0.009$ & $0.286 \pm 0.008$ \\
\bottomrule
\end{tabular}
\end{table}

Table~\ref{tab:alpha_both} summarizes the fitted coefficients at both energies. The Plan~B method successfully recovers the input $\alpha$ in both cases. At $\sqrt{s} = 4.8$~GeV, where the angular modulation is significant, the Plan~A method underestimates $\alpha$ by roughly $34\%$, while the Plan~B method restores the correct value within one standard deviation.

We note that at $\sqrt{s}=3.686$~GeV, the Plan~B result ($\alpha = 0.034 \pm 0.003$) has a statistical uncertainty comparable to that of the true-only result ($\alpha = 0.037 \pm 0.003$). This is a natural consequence of the weighting scheme: Plan~B uses both solutions per event, so the effective statistical power is larger than in the true-only analysis, which discards the spurious solution, and the weighted histogram benefits from partial cancellation of bin-to-bin Poisson fluctuations when both solutions contribute. The gain is modest at this energy, where the two solutions are close to each other, and is larger at $\sqrt{s}=4.8$~GeV. The agreement between $\alpha_{\rm PlanB}$ and $\alpha_{\rm input}$ at both energies is within about one standard deviation, confirming that the method is statistically consistent.

\subsection{Iterative convergence}

The fixed-point nature of the Plan~B method is demonstrated by the iterative procedure. We test the convergence from a wide range of initial guesses $\alpha^{(0)}$ at $\sqrt{s} = 4.8$~GeV. In all cases, the iteration converges to the vicinity of the true $\alpha \approx 0.29$ within 5--8 steps, as shown in Table~\ref{tab:iteration}:

\begin{table}[H]
\centering
\caption{Iterative convergence of the Plan~B method at $\sqrt{s} = 4.8$~GeV for different initial guesses $\alpha^{(0)}$. Convergence to $\alpha \approx 0.29$ is achieved from all tested starting values. Empty cells indicate that the iteration had already satisfied $|\alpha^{(n+1)} - \alpha^{(n)}| < \delta$ at that step, so the iteration was terminated.}
\label{tab:iteration}
\begin{tabular}{cccccc}
\toprule
Iteration $n$ & \multicolumn{5}{c}{$\alpha_{\rm fit}^{(n)}$ starting from $\alpha^{(0)}$} \\
\cmidrule(lr){2-6}
& $0.000$ & $0.200$ & $0.500$ & $0.700$ & $0.900$ \\
\midrule
0 & 0.000 & 0.200 & 0.500 & 0.700 & 0.900 \\
1 & 0.115 & 0.263 & 0.352 & 0.372 & 0.388 \\
2 & 0.195 & 0.283 & 0.301 & 0.308 & 0.312 \\
3 & 0.248 & 0.289 & 0.293 & 0.294 & 0.295 \\
4 & 0.272 & 0.290 & 0.291 & 0.291 & 0.291 \\
5 & 0.283 & 0.291 & 0.291 & 0.291 & 0.291 \\
6 & 0.288 & 0.291 & & & \\
7 & 0.290 & & & & \\
\bottomrule
\end{tabular}
\end{table}

\subsection{Sensitivity to Experimental Effects}
\label{sec:experimental}

In a realistic experimental environment, several effects can modify the observed $\cos\theta$ distribution relative to the underlying production distribution $f(\cos\theta)$:

\begin{enumerate}
\item {\bf Beam energy spread.} At BEPC-II, the center-of-mass energy has a Gaussian spread of $\sigma_{E} \approx 1.2$~MeV. Since the angular coefficient $\alpha$ depends on $\sqrt{s}$ through Eq.~(\ref{eq:alpha_beta}), this spread introduces an effective smearing of $\alpha$. However, the dependence is weak: at $\sqrt{s}=3.686$~GeV, $\partial\alpha/\partial\sqrt{s} \approx 0.27$~GeV$^{-1}$, so that a spread of $\sigma_{E} \approx 1.2$~MeV shifts the effective $\alpha$ by only $\approx 3\times10^{-4}$, which is negligible.

\item {\bf Detector resolution.} The tracking momentum resolution ($\delta p/p \approx 0.5\%$) smears the measured kaon momenta. This propagates into the reconstructed $\cos\theta$ values through the constraint equations~(\ref{eq:c_params}) and~(\ref{eq:linear_system}), broadening the observed angular distribution. An efficiency function $\eta(\cos\theta)$ describing the detector acceptance and reconstruction efficiency may also introduce a non-uniform modulation of the observed distribution.

\item {\bf Generalization of the method.} When detector effects are present, the observed angular distribution is not $f(\cos\theta)$ but an effective distribution $f_{\rm eff}(\cos\theta) = f(\cos\theta) \cdot \eta(\cos\theta) \otimes R(\cos\theta)$, where $\eta(\cos\theta)$ is the detection efficiency and $R(\cos\theta)$ represents resolution smearing. The iterative method can be extended to this case by replacing the weight function $f(\cos\theta;\alpha)$ with the convolution of $f(\cos\theta;\alpha)$ with the detector response. The fixed-point structure of the iteration is preserved as long as the detailed balance condition remains approximately valid for the effective distribution. A full study including realistic detector simulation is deferred to future work.
\end{enumerate}

\subsection{Degenerate case: flat angular distribution}

We note a special degenerate case of the Plan~B method. When the true angular distribution is flat ($\alpha = 0$, i.e., $f(\cos\theta) = \text{const}$), the spurious solution is also uniformly distributed, and the weights $w_T = w_S = 1/2$ produce a flat weighted histogram regardless of the value of $\alpha$ used in the weight function. In this limit, the iteration trivially converges to $\alpha = 0$ from any starting value---but this convergence provides no evidence that the method has recovered the true distribution, since {\it any} input $\alpha$ would produce the same flat output. For practical analyses, if the extracted $\alpha$ is consistent with zero within uncertainty, one should verify using independent methods whether the underlying distribution is genuinely flat. This degenerate case does not affect the method's validity for distributions with significant modulation ($\alpha \gtrsim 0.1$), which is the regime where Plan~A fails and Plan~B is most needed.

\section{Summary and Remarks}
\label{sec:summary}

We have presented a self-consistent weighting method for handling the two-fold ambiguity in $\tau$ momentum reconstruction. We derived analytic expressions for the $\tau$ momentum direction from the kinematic constraints, recovering the two-fold ambiguity in closed form. We showed that the Plan~A treatment systematically biases the extracted angular coefficient whenever the underlying distribution is not uniform, because the spurious solution tends toward a nearly uniform distribution; the bias grows with the modulation and is resolved at both energies studied here. We proposed a weighted method (Plan~B) that assigns normalized weights $w_{i} = f(\cos\theta_{i})/[f(\cos\theta_{1})+f(\cos\theta_{2})]$ to each solution, and proved a fixed-point theorem showing that $h(x) = f(x)[1+\varepsilon(x)]$ with $\varepsilon(x) \ll 1$ under the detailed balance condition. Monte Carlo simulations at $\sqrt{s} = 3.686$~GeV and $4.8$~GeV validate the method. At $4.8$~GeV ($\alpha \approx 0.292$), Plan~B recovers $\alpha = 0.286 \pm 0.008$, while the Plan~A method yields $\alpha = 0.193 \pm 0.009$. The iteration converges to the true $\alpha$ within 5--8 steps from a wide range of initial guesses, demonstrating the robustness of the fixed-point method.

The apparent success of the Plan~A method at the $\psip$ threshold~\cite{Zhou2026} is explained by the near-uniformity of the QED angular distribution ($\alpha \approx 0.037$), where the spurious solution is almost uniform as well and the resulting bias is small in absolute terms. That bias is nevertheless resolved once the statistical precision is sufficient (Sec.~IV.B), whereas the present method removes it and remains applicable to distributions with arbitrary modulation. The method does not require event-by-event discrimination and operates entirely at the statistical level, making it suitable for experimental analyses where vertex information is insufficient. Future work should extend the method to include the full spin correlation information of the $\tau$ pair, and to other decay channels such as $\tau \to \pi\nu_{\tau}$ and $\tau \to \rho\nu_{\tau}$.

\section*{Acknowledgments}

This work is supported by the National Natural Science Foundation of China under Grant No. 12575112, and the National Key Research and Development Program of China under Contract No. 2025YFA1613900.

\appendix

\section{Proof of the Fixed-Point Theorem}
\label{sec:appendix_proof}

We present the detailed proof of Theorem~1 stated in Section~\ref{sec:method}.

\subsection*{Setup}

Let $T \sim f(t)$ be the true solution and $S \sim q(s \mid T)$ be the spurious solution, where $f(t)$ is the probability density function of the true angular variable and $q(s \mid t)$ is the conditional density of the spurious solution given the true solution. The Plan~B weighted density is defined as
\begin{equation}
h(z) = \mathbb{E}_{T\sim f,\, S\sim q(\cdot\mid T)}\left[\delta(z-T)\frac{f(T)}{f(T)+f(S)} + \delta(z-S)\frac{f(S)}{f(T)+f(S)}\right].
\label{eq:A1}
\end{equation}

\subsection*{Expansion of the expectation}

Expanding the expectation over the joint distribution $p(t,s) = f(t)q(s \mid t)$:
\begin{equation}
\begin{aligned}
h(z) &= \iint f(t)q(s \mid t) \left[\delta(z-t)\frac{f(t)}{f(t)+f(s)} + \delta(z-s)\frac{f(s)}{f(t)+f(s)}\right] dt\,ds \\[4pt]
&= \underbrace{\iint f(t)q(s \mid t)\,\delta(z-t)\frac{f(t)}{f(t)+f(s)} dt\,ds}_{I_{1}} \;+\; \underbrace{\iint f(t)q(s \mid t)\,\delta(z-s)\frac{f(s)}{f(t)+f(s)} dt\,ds}_{I_{2}}.
\end{aligned}
\label{eq:A2}
\end{equation}

\subsection*{Evaluation of $I_{1}$}

The delta function $\delta(z-t)$ fixes $t = z$:
\begin{equation}
I_{1} = f(z)^{2} \int \frac{q(s \mid z)}{f(z) + f(s)} \, ds.
\label{eq:A3}
\end{equation}

\subsection*{Evaluation of $I_{2}$}

The delta function $\delta(z-s)$ fixes $s = z$:
\begin{equation}
I_{2} = f(z) \int \frac{f(t) \, q(z \mid t)}{f(t) + f(z)} \, dt.
\label{eq:A4}
\end{equation}

\subsection*{Application of detailed balance}

The detailed balance condition states that
\begin{equation}
f(t) \, q(z \mid t) = f(z) \, q(t \mid z) \qquad \forall\, t, z.
\label{eq:A5}
\end{equation}

Substituting Eq.~(\ref{eq:A5}) into Eq.~(\ref{eq:A4}):
\begin{equation}
I_{2} = f(z) \int \frac{f(z) \, q(t \mid z)}{f(t) + f(z)} \, dt = f(z)^{2} \int \frac{q(t \mid z)}{f(t) + f(z)} \, dt.
\label{eq:A6}
\end{equation}

\subsection*{Combining $I_{1}$ and $I_{2}$}

Since the integration variables $s$ in $I_{1}$ and $t$ in $I_{2}$ are dummy, we can rename them to a common variable $y$:
\begin{equation}
h(z) = I_{1} + I_{2} = 2 f(z)^{2} \int \frac{q(y \mid z)}{f(z) + f(y)} \, dy.
\label{eq:A7}
\end{equation}

\subsection*{Analysis of the integral}

Define the integral factor
\begin{equation}
I(z) = 2 f(z) \int \frac{q(y \mid z)}{f(z) + f(y)} \, dy,
\label{eq:A8}
\end{equation}
so that $h(z) = f(z) \cdot I(z)$. Using the algebraic identity
\begin{equation}
\frac{2f(z)}{f(z) + f(y)} = 1 + \frac{f(z) - f(y)}{f(z) + f(y)},
\label{eq:A9}
\end{equation}
we obtain
\begin{equation}
\begin{aligned}
I(z) &= \int q(y \mid z) \left[1 + \frac{f(z) - f(y)}{f(z) + f(y)}\right] dy \\[4pt]
&= \int q(y \mid z) \, dy + \int q(y \mid z) \, \frac{f(z) - f(y)}{f(z) + f(y)} \, dy \\[4pt]
&= 1 + \varepsilon(z),
\end{aligned}
\label{eq:A10}
\end{equation}
where we used the normalization $\int q(y \mid z) dy = 1$ and defined the correction term
\begin{equation}
\varepsilon(z) = \int q(y \mid z) \cdot \frac{f(z) - f(y)}{f(z) + f(y)} \, dy.
\label{eq:A11}
\end{equation}

\subsection*{Final result}

Substituting back into $h(z) = f(z) \cdot I(z)$:
\begin{equation}
h(z) = f(z) \cdot \big[1 + \varepsilon(z)\big].
\label{eq:A12}
\end{equation}

\subsection*{Bound on the correction term}

For $f(z) = (1 + \alpha z^{2})/N$ with $\alpha \in [0, 1]$ on $z \in [-1, 1]$, the function varies between $f_{\min} = 1/N$ and $f_{\max} = (1+\alpha)/N$. The maximum relative difference is
\begin{equation}
\left|\frac{f(z) - f(y)}{f(z) + f(y)}\right| \leq \frac{\alpha}{2 + \alpha} \leq \frac{1}{3} \quad (\text{for } \alpha \leq 1).
\label{eq:A13}
\end{equation}
Since $q(y \mid z)$ is a probability density, the correction term is bounded by $|\varepsilon(z)| \leq \alpha/(2+\alpha)$. For $\alpha = 0.7$, $|\varepsilon| \leq 0.26$, and in practice the integral averages to much smaller values ($\sim 0.01$--$0.02$) due to cancellations between positive and negative contributions.

This completes the proof. $\square$

\section{Verification of the Detailed Balance Condition}
\label{sec:appendix_db}

\begin{figure}[H]
\centering
\includegraphics[width=0.8\textwidth]{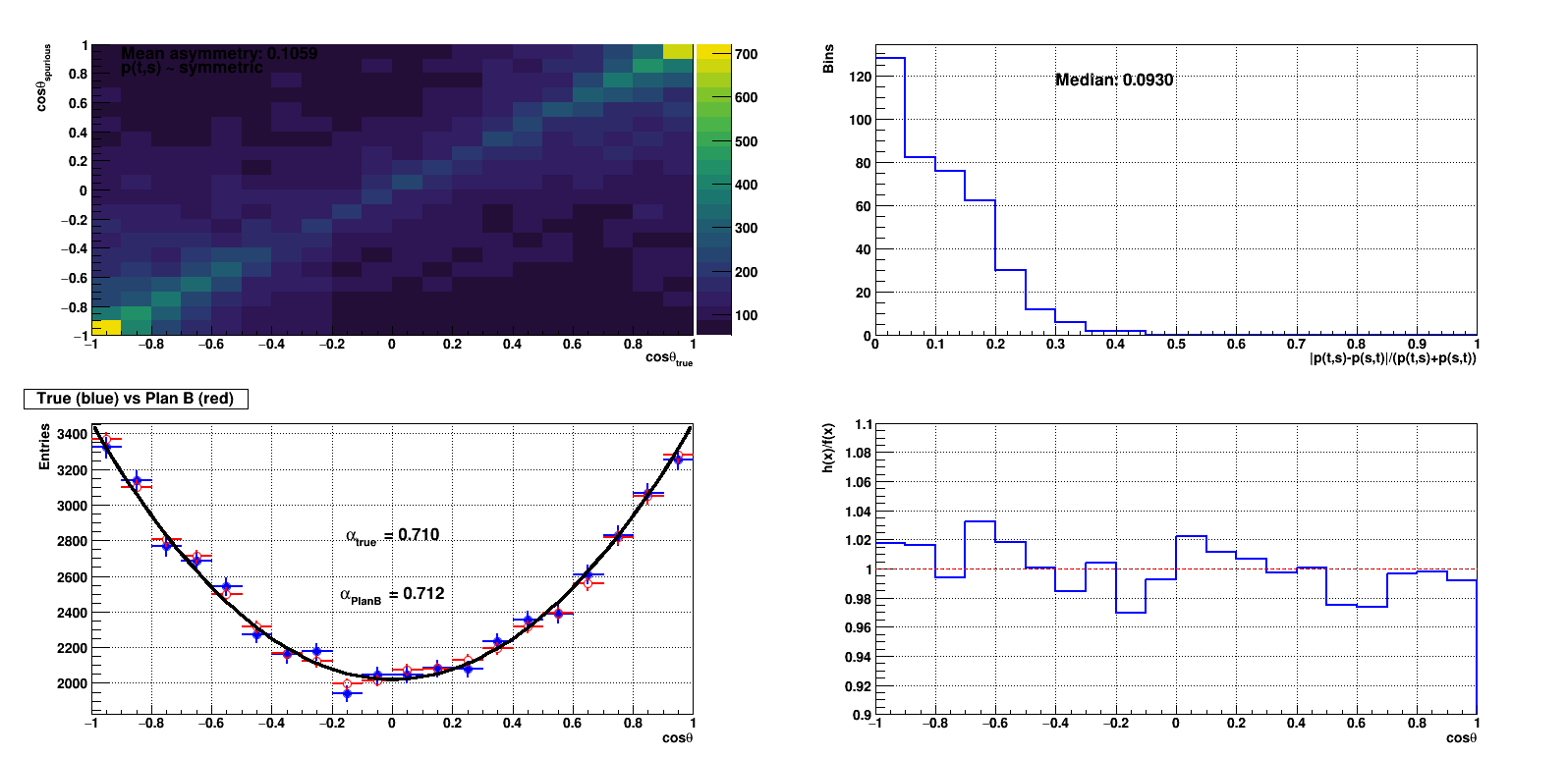}
\caption{Verification of the detailed balance condition: (a) the joint distribution $p(t,s)$ of true and spurious $\cos\theta$ values; (b) the true (blue) and Plan~B weighted (red) $\cos\theta$ distributions, with fits giving $\alpha_{\rm true} = 0.710$ and $\alpha_{\rm PlanB} = 0.712$ (input $\alpha = 0.7$); (c) the distribution of the asymmetry $|p(t,s)-p(s,t)|/(p(t,s)+p(s,t))$, whose median is $0.093$; and (d) the ratio $h(x)/f(x)$ of the Plan~B weighted to the true distribution, within $3\%$ of unity.}
\label{fig:detailed_balance}
\end{figure}

Figure~\ref{fig:detailed_balance} shows the joint distribution $p(t,s) = f(t)q(s \mid t)$ of the true and spurious $\cos\theta$ values, together with the asymmetry measure $|p(t,s) - p(s,t)|/(p(t,s)+p(s,t))$. The mean asymmetry is approximately $0.11$, and the median is $0.09$, confirming that the detailed balance condition holds to good approximation.

\begin{thebibliography}{99}

\bibitem{PDG2024}
F.~Takahashi {\it et al.} (Particle Data Group), Int. J. Mod. Phys. A 41, 2630011 (2026).

\bibitem{BESIII2020}
M.~Ablikim {\it et al.} (BESIII Collaboration),
Chin.\ Phys.\ C {\bf 44}, 040001 (2020).

\bibitem{BESIII2014}
M.~Ablikim {\it et al.} (BESIII Collaboration),
Phys.\ Rev.\ D {\bf 90}, 012001 (2014).

\bibitem{Fabbrichesi2024}
M.~Fabbrichesi and L.~Marzola,
Phys.\ Rev.\ D {\bf 110}, 076004 (2024).

\bibitem{Han2025}
T.~Han, M.~Low, and Y.~Su,
JHEP {\bf 10}, 217 (2025).

\bibitem{Cheng2025}
K.~Cheng, T.~Han, and M.~Low,
Phys.\ Lett.\ B {\bf 868}, 139675 (2025).

\bibitem{Kuhn1993}
J.~H.~K\"uhn,
Phys.\ Lett.\ B {\bf 313}, 458 (1993).

\bibitem{Zhou2026}
X.~Zhou, J.~Zhang, X.~Wan, Y.~Wang, and X.~Mo,
arXiv:2605.26724 [hep-ph] (2026).

\bibitem{Ehat2024}
K.~Ehat\"aht, M.~Fabbrichesi, L.~Marzola, and C.~Veelken,
Phys.\ Rev.\ D {\bf 109}, 032005 (2024).

\end{thebibliography}
\end{document}